\documentclass[12pt,a4paper]{article}
\usepackage{mathrsfs}
\usepackage{amsfonts}
\usepackage[intlimits]{amsmath}
\usepackage{epsf, cite}

\def\lmatrix{\left(\begin{array}}
\def\rmatrix{\end{array}\right)}
\def\bea{\begin{eqnarray}}
\def\eea{\end{eqnarray}}
\def\nn{\nonumber}
\def\half{\frac{1}{2}}
\def\tr{{\rm Tr\,}}
\def\det{{\rm det\,}}
\def\diag{{\rm diag\,}}

\def\R{{\mathbb R}}
\def\C{{\mathbb C}}
\def\vesszo{{\,\prime}}
\def\curly{\mathscr}
\def\dash{\textendash\, }

\newenvironment{narrowmargin}{
\begin{list}{}
    {
    \setlength{\parindent}{0pt}
    \setlength{\topsep}{0pt}
    \setlength{\leftmargin}{1cm}
    \setlength{\rightmargin}{1cm}
    \setlength{\listparindent}{\parindent}
    \setlength{\itemindent}{\parindent}
    \setlength{\parsep}{\parskip}
    }\item[]}
{\end{list}}

\newcommand{\sn}[2]{sn(#1\,|\,#2)}
\newcommand{\cn}[2]{cn(#1\,|\,#2)}
\newcommand{\dn}[2]{dn(#1\,|\,#2)}
\newcommand{\snn}[2]{sn^2(#1\,|\,#2)}

\newcommand{\dnn}[2]{dn^2(#1\,|\,#2)}
\newcommand{\fejezet}[1]{\bigskip\noindent{\it\large #1}\bigskip}

\begin{document}


\begin{center}{\Large\bf M2-branes stretching between M5-branes}\\

\bigskip\bigskip
D\'aniel N\'ogr\'adi\\
\bigskip

{\em Institute Lorentz for Theoretical Physics, University of Leiden\\
P.O.Box 9506, 2300 RA Leiden, The Netherlands}

\smallskip
and
\smallskip

{\em Department of Theoretical Physics, University of Wuppertal\\
Gauss 20, 42119 Wuppertal, Germany}
\bigskip

\bigskip
{\tt\footnotesize nogradi@lorentz.leidenuniv.nl}

\end{center}

\bigskip\bigskip

\begin{narrowmargin}
A generalization of Nahm's equation has been recently conjectured by Basu and Harvey to be
the BPS condition describing the bound state of a stack of M2-branes ending on an M5-brane.
In this note exact solutions are presented for the proposed BPS equation \dash which
is from the point of view of the M2-brane world-volume dynamics \dash with boundary
conditions appropriate for M2-branes stretching between two M5-branes.
Unfortunately, since the action for multiple M5-branes or for multiple coincident M2-branes
is not known, one can only resort to consistency checks of the proposal instead of a direct comparison
of the M2 and M5 world-volume point of views. The existence of our solutions
should be seen as such a consistency check of the conjecture, and also as a source of
new insight into the dynamics of multiple M2 and M5-branes.
\end{narrowmargin}

\newpage

\section{Introduction}

In this work a specific configuration of M2-branes and M5-branes is studied in M-theory. In principle
there would be (at least) three different approaches to such a problem. One could study the 11-dimensional localized supergravity
solutions, the world-volume dynamics of the M2-branes or the world-volume dynamics of the M5-branes.
In our example \dash a stack of M2-branes stretching between two separated M5-branes \dash the 
localized supergravity solution seems to be very difficult to find, the 3-dimensional multiple M2-brane world-volume action
is not known, nor the 6-dimensional action for two separated M5-branes thus it seems that the possibilities
are rather limited.

Although the action for multiple M2-branes is not known, recently a proposal has been made by Basu
and Harvey for the BPS condition describing the bound state of a stack of M2-branes ending on an M5 \cite{Basu:2004ed}.
Their equation is from the world-volume point of view of the M2-branes and it is
analogous to the BPS condition describing the bound state of a stack of D1-branes
ending on a D3-brane in type IIB string theory. More precisely, the BPS condition for the world-sheet
dynamics of the D1-branes is Nahm's equation \cite{Nahm:1979yw, Nahm, Diaconescu:1996rk}
and the Basu-Harvey conjecture states
that an appropriate generalization of the latter may be interpreted as a BPS condition for the
M2-M5 brane system.

In this note exact and rather explicit solutions for the generalized Nahm equation are found for a specific geometry, namely a stack of
M2-branes stretching between two separated M5-branes.

Related \dash and mostly simpler \dash brane configurations have been
studied extensively of course in all three aspects mentioned above. As for supergravity, M2 and M5-brane solutions
were found long ago \cite{Duff:1990xz, Gueven:1992hh} as well as orthogonally intersecting M2 or M5-branes
\cite{Papadopoulos:1996uq}. There also exist solutions for a single M2 intersecting an M5 \cite{Tseytlin:1996bh, Gauntlett:1996pb}.
Furthermore, branes may end on other branes \cite{Strominger:1995ac, Townsend:1996em} and an M2 may stretch between two separated M5's
\cite{Tseytlin:1997cs}.  A thorough review of these matters with an emphasis on intersecting brane solutions
of 11 and 10 dimensional supergravity can be found in  \cite{Gauntlett:1997cv}. However, it might
be worth reminding that these intersecting brane solutions are such that the branes are not fully
localized.

The world-volume dynamics of branes have also been investigated in a variety of cases. The membrane
action has been given in \cite{Bergshoeff:1987cm}. For a stack of membranes ending on an M5,
although the world-volume action for multiple membranes is not known, the solution on the world-volume
of the M5 has been found and is the self-dual string \cite{Howe:1997ue}. Unfortunately not only the
action for multiple coincident membranes is not known, so is the case for separated M5-branes.
This is in contrast to the situation in string theory where the world-volume dynamics of
D-branes is given by well-understood gauge theories.
For instance, if coincident D1-branes stretch
between separated D3's, then the corresponding BPS condition is the Nahm equation from the point of
view of the D1's and the Bogomolny equation from the point of view of the D3's \cite{Diaconescu:1996rk}.

Guided by the analogy to the well-understood D1-D3 brane system 
\cite{Callan:1997kz,Gibbons:1997xz,Constable:1999ac}
the conjecture of Basu and Harvey could be made without
knowing the action for a stack of membranes ending on an M5. The descriptions for this bound state by
the proposed generalization of Nahm's equation on the one hand (membranes point of view) and the black string solution on the other hand
(M5 point of view) were shown to match for some relevant quantities \cite{Basu:2004ed}.
The fact that exact solutions \dash presented in this short note \dash exist for a stack of membranes stretching between two
separated M5-branes should be seen as a further consistency check.

It is also possible to modify the Basu-Harvey equation in a way that describes membranes ending on intersecting
M5-branes which fact may also be viewed as an additional consistency check \cite{Berman:2005re}.

\section{Membranes ending on M5-branes}

\fejezet{Generalized Nahm equation}

An equation due to Basu and Harvey, that is conjectured to describe a stack of $N$ membranes ending on an M5-brane, is
\footnote{Note the difference in the coefficient of the second term relative to the first version of \cite{Basu:2004ed}.
I thank Jeff Harvey for pointing out this small mistake there.}
\bea
\label{hb}
\frac{dX_i}{ds} + \frac{M_{11}^3}{8\pi\sqrt{2N}} \frac{1}{4!}\,\varepsilon_{ijkl} [ G_5,X_j,X_k,X_l ] = 0\,.
\eea
Here $M_{11}$ is the 11-dimensional Planck mass and the four variables $X_i(s)$ are $N\times N$ complex matrices \cite{Basu:2004ed}.
The quantum Nambu bracket \cite{Nambu:1973qe} for any four matrices $A_i$ is defined by
\bea
~[A_1,A_2,A_3,A_4] = \varepsilon_{ijkl} A_i A_j A_k A_l\,,\nn
\eea
and the $N\times N$ matrix $G_5$ is the difference of two projections associated with two $Spin(4)$ representations,
see \cite{Basu:2004ed} for details.
Essentially it is a higher dimensional analog of the (Euclidean) Dirac matrix
$\gamma_5=\frac{1}{4!}\varepsilon_{ijkl}\, \gamma_i\, \gamma_j\, \gamma_k\, \gamma_l$ and in this note we will
restrict ourselves to $N=4$, in which case $G_5=\gamma_5$. For an earlier appearance of the quantum Nambu bracket
along with $G_5$ in a stringy context see \cite{Sheikh-Jabbari:2004ik}, and for a physical interpretation
of the ``chirality'' matrix $G_5$ the reader is referred to \cite{Sheikh-Jabbari:2005mf}.
The parameter $s$ runs orthogonal to the M5 and along the M2's.

It is well-known that a bound state of $N$ D1-branes ending on a D3-brane 
in IIB string theory may be described \dash from the point of view of the D1 world-sheet dynamics \dash by Nahm's equation
\cite{Nahm:1979yw, Nahm, Diaconescu:1996rk}. The proposed generalization for M-theory that ought to describe a stack
of membranes ending on an M5 is equation (\ref{hb}), assuming for simplicity dependence on one spatial dimension only.
Thus (\ref{hb}) should be understood as a  world-volume equation for the membranes, more precisely, it
should be the BPS equation describing the bound state. However, the action for coincident membranes from which
this equation would follow as a BPS condition, is not known and the Basu-Harvey proposal remains a conjecture,
although well-motivated. 

We will assume $\{\gamma_5,X_i\} = 0$, thus (\ref{hb}) can be written
\bea
\label{hb2}
\frac{dX_i}{ds} + \frac{1}{12}\,\varepsilon_{ijkl}\, \gamma_5 X_j X_k X_l = 0\,,
\eea
with an appropriate choice of units.

The range of $s$ depends on the configuration of the branes. If a single M5 is placed at $s=0$ and the
membranes that end on it extend to infinity, then the range of $s$
is the semi-infinite interval $(0,+\infty)$. The boundary condition is then the following,
\bea
\label{bc0}
X_i(s) = \frac{i\gamma_i}{\sqrt{s}} \qquad {\rm for}\quad s\to 0\,,
\eea
which is the only power-like behaviour compatible with (\ref{hb2}).
In fact, the above form is a solution of (\ref{hb2}) on the whole range $(0,+\infty)$.

If a stack of four membranes are suspended between two M5-branes, placed
at $s=\pm s_0$, then the range of $s$ is the finite open interval $(-s_0,s_0)$. The natural boundary conditions 
in this case are
\bea
\label{bc11}
X_i(s) = \frac{i\gamma_i}{\sqrt{s_0\mp s}} +\;\ldots \qquad {\rm for}\quad s\to\pm s_0\,.
\eea
In this note we seek solutions of (\ref{hb2}) on $(-s_0,s_0)$ with the above boundary conditions.

Unfortunately, the action for multiple M5-branes or multiple coincident membranes, either of which would be necessary to compare our results
with something, is not known. Thus the only justification for identifying our solutions with a specific
M2-M5 geometry
are the consistency checks performed in \cite{Basu:2004ed} and \cite{Berman:2005re} supporting equation (\ref{hb}) as a viable candidate
for describing such a bound state. 

\fejezet{Exact solutions}

In order to obtain explicit and exact solutions of (\ref{hb2}) with boundary conditions (\ref{bc11})
we pick the ansatz $X_i = i a_{ij}\,\gamma_j$, with a $4\times 4$ matrix $a=(a_{ij})$.
The generalized Nahm equation (\ref{hb2}) gives for $a_{ij}$,
\bea
\label{dx}
\frac{\,da_{ij}}{\!ds} + \frac{1}{12}\,\varepsilon_{iklm}\,\varepsilon_{jnpq}\,a_{kn}\, a_{lp}\, a_{mq} = 0\,. 
\eea
Now using the formula $\varepsilon_{ijkl}\, \det a = \varepsilon_{mnpq}\, a_{im}\, a_{jn}\, a_{kp}\, a_{lq}$, for
the determinant of a $4\times 4$ matrix, it follows that (suppressing indices $i,j,\ldots$, etc.) 
\bea
\label{adet}
\frac{d(aa^T)}{\!ds} + \det a =0\,,
\eea
with $a^T$ meaning the transpose of $a$.
Thus the traceless part of the symmetric matrix $aa^T$ is conserved. Let us split $aa^T$ into a scalar
part and a traceless symmetric part by writing $aa^T = x - C$, where $C=(C_{ij})$ is a constant,
symmetric and traceless matrix and $x = \tr(aa^T)/4$ is an unknown function of $s$. Then (\ref{adet})
implies
\bea
\label{df}
\left(\frac{dx}{ds}\right)^2 = (x-x_1)(x-x_2)(x-x_3)(x-x_4)\,,
\eea
where the $x_i$ are the eigenvalues of $C$, hence $x_1+x_2+x_3+x_4=0$. Thus we are led to study the
elliptic curve
\bea
\label{curve}
y^2 = (x-x_1)(x-x_2)(x-x_3)(x-x_4)\,.
\eea

It is known on general grounds that solutions of (\ref{df}) are elliptic functions with periods determined
by the roots $x_i$ \cite{whittaker}. Concretely, let us set
\bea
\label{param}
m&=& \frac{(x_2-x_3)(x_1-x_4)}{(x_2-x_4)(x_1-x_3)}\nn\\ 
D&=& \half \sqrt{(x_2-x_4)(x_1-x_3)}\,,
\eea
and further impose
\bea
\label{sn}
sn^2(Ds_0|m)&=&\frac{x_3-x_1}{x_3-x_2}\,,
\eea
where $sn(z|m)$ and subsequently $cn(z|m)$ and $dn(z|m)$ are the Jacobi elliptic functions.
It will not lead to confusion to omit the argument $m$ and we will simply write $sn(z), cn(z)$ and $dn(z)$. In
addition we introduce the useful function
\bea
\label{phi}
\phi(s) = \frac{2D \,sn(Ds_0)\, cn(Ds_0)\, dn(Ds_0)}{sn^2(Ds_0)-sn^2(Ds)}\,.
\eea
Then we have the following exact solution of equation (\ref{df}) that may be checked by direct
substitution or by uniformization of the curve (\ref{curve}),
\bea
\label{phisolutions}
x(s)-x_1 &=& \phi(s)\nn\\
x(s)-x_2 &=& \phi(s)\,\frac{sn^2(Ds)}{sn^2(Ds_0)}\nn\\
x(s)-x_3 &=& \phi(s)\,\frac{cn^2(Ds)}{cn^2(Ds_0)}\\
x(s)-x_4 &=& \phi(s)\,\frac{dn^2(Ds)}{dn^2(Ds_0)}\nn\,.
\eea
Naturally, any of these expressions define $x(s)$ unambiguously, but we will see that it is convenient
to have explicit forms for $x(s)-x_i$ for all $i$. The definitions and a number of useful identities
involving the Jacobi and Weierstrass elliptic functions as well as generalities on the curve (\ref{curve})
and its uniformization are given in an appendix.

As advertised, $x(s)$ is an elliptic function indeed and has a simple pole at the location
of the M5-branes,
\bea
\label{pole}
x(s) = \frac{1}{s_0\mp s} + \ldots \qquad {\rm for}\quad s\to\pm s_0\,,
\eea
which can be read off easily from (\ref{phi}). For generic complex values of $D$ and $m$ there
are no other poles on the interval $(-s_0,s_0)$, but if they were both real one needed to impose
$Ds_0 < K(m)$. Here $K(m)$ is the complete elliptic integral of the first kind, see the appendix for its definition. 

Let us proceed by reconstructing $X_i(s)$ from $x(s)$. It may be assumed
that the constant, symmetric and traceless matrix $C$ is diagonal.
Since $aa^T=x-C$, $aa^T$ is diagonal as well and $a$ can be chosen to be
\bea
\label{asol}
a(s) = \diag(\sqrt{x(s)-x_1},\sqrt{x(s)-x_2},\sqrt{x(s)-x_3},\sqrt{x(s)-x_4})\,.
\eea
Every other possible choice may be obtained by multiplying the above form by an orthogonal matrix from the right, since
such a transformation does not affect $aa^T$ and consequently nor $x$. However, it does affect the boundary
condition (\ref{bc11}) and hence we ignore these extra moduli.

Thus, using the ansatz $X_i = ia_{ij}\gamma_j$ we conclude that the most general solution (within our ansatz)
of the generalized Nahm equation is the 2-parameter family 
\bea
\label{solution}
X_1(s) &=& i\sqrt{\phi(s)}\,\gamma_1 \nn\\
X_2(s) &=& i\sqrt{\phi(s)}\, \frac{sn(Ds)}{sn(Ds_0)}\,\gamma_2 \nn\\
X_3(s) &=& i\sqrt{\phi(s)}\, \frac{cn(Ds)}{cn(Ds_0)}\,\gamma_3 \\
X_4(s) &=& i\sqrt{\phi(s)}\, \frac{dn(Ds)}{dn(Ds_0)}\,\gamma_4\,, \nn
\eea
parameterized by ($D, m$) up to $O(4)$ where $U\in O(4)$ acts according to $a\to Ua\,U^T$. This transformation
corresponds to having a non-diagonal $C$. The pole
(\ref{pole}) shows that the boundary condition (\ref{bc11}) is fulfilled and the constraint (\ref{sn}) was in fact necessary
for having the poles in $x(s)$ at the prescribed points $s=\pm s_0$. It might also be worth noting that the above solutions
bare some resemblance to the ones found for $SU(2)$ monopoles or equivalently for the
D1-D3 brane system \cite{Brown:1982gz, Dancer:1992kn}; see also \cite{Papageorgakis:2005zf}.

Thus, assuming real $D$ and $m$ parameters as well as rotations by $O(4,\R)$ as opposed to complex $(D,m)$ moduli
and $O(4,\C)$, we have obtained an 8 real dimensional relative moduli space. Relative, because there is an obvious
symmetry $X_i \to X_i + v_i$ with an arbitrary 4-vector $v_i$ corresponding to over-all translations which was not
taken into account.

It is easy to recover the single M5-brane solution (\ref{bc0}) by shifting the M5's to $s=0$ and
$s=2s_0$, and then taking the limit $s_0\to\infty$. Let us suppose that both $D$ and $m$ are real,
then in order to maintain $s_0 D < K(m)$ for fixed $m$, one needs to take
the $D\to 0$ limit first, followed by pushing one of the M5's to infinity, eventually leading to $X_i = i\gamma_i/\sqrt{s}$, as it should.

\fejezet{Coincident roots}

We have so far assumed that all roots were different, but the above general solution simplifies considerably if the elliptic
curve (\ref{curve}) degenerates. First, let us assume that two roots coincide, $x_4=x_1$. In this case $m=0$ and the Jacobi
functions reduce to trigonometric (or hyperbolic) functions. Explicitly, we have
\bea
\label{solutiontri}
\phi(s)\!\!\!&=&\!\!\!\frac{D\sin(2Ds_0)}{\sin(D(s_0-s)) \sin (D(s_0+s))}\nn\\
X_1(s) = i\sqrt{\phi(s)}\,\gamma_1&\qquad& X_2(s) = i\sqrt{\phi(s)}\,\frac{\sin(Ds)}{\sin(Ds_0)}\,\gamma_2\\
X_4(s) = i\sqrt{\phi(s)}\,\gamma_4&\qquad& X_3(s) = i\sqrt{\phi(s)}\,\frac{\cos(Ds)}{\cos(Ds_0)}\,\gamma_3\,.\nn
\eea
Note that for both real and purely imaginary $D$, $\phi(s)$ is positive on the interval $(-s_0,s_0)$.

If a further degeneracy occurs and $x_3=x_2=-iD$ as well as $x_4=x_1=iD$ then the solutions will
have a pole, provided $D$ is real. Let us assume this, then the condition analogous to (\ref{sn}) gives
$s_0 D=\pi/2$ and we find
\bea
\label{solutiontri2}
\phi(s)\!\!\! &=&\!\!\!\frac{i\pi}{2s_0} \frac{\exp\left(\frac{i\pi s}{2s_0}\right)}{\cos(\frac{\pi s}{2s_0})}\\
X_1(s) = i\sqrt{\phi(s)}\,\gamma_1\qquad X_2(s) = i\sqrt{\bar\phi(s)}\,\gamma_2\!\!\!&\qquad&\!\!\!X_3(s) = i\sqrt{\bar\phi(s)}\,\gamma_3\qquad
X_4(s) = i\sqrt{\phi(s)}\,\gamma_4\,.\nn
\eea

Finally, we consider the case of three coinciding roots, $x_1=-3D$ and $x_2=x_3=x_4=D$. Again, $D$ has to be real and in order to have the poles
at $\pm s_0$ one has to impose $s_0 D = 1/2$, which leads to the simple solutions
\bea
\label{solutionsima}
\phi(s)\!\!\!&=&\!\!\! \frac{2s_0}{(s_0-s)(s_0+s)}\\
X_1(s) = i\frac{s}{s_0}\, \sqrt{\phi(s)}\,\gamma_1 \qquad
X_2(s) = i\sqrt{\phi(s)}\,\gamma_2\!\!\!&\qquad&\!\!\!
X_3(s) = i\sqrt{\phi(s)}\,\gamma_3\qquad
X_4(s) = i\sqrt{\phi(s)}\,\gamma_4\,.\nn
\eea
Just as in (\ref{solutiontri}), $\phi(s)$ is positive between the two M5-branes. The above form is in fact
the $D\to 0$ limit of the most general solution (\ref{solution}) and the single M5-brane solution (\ref{bc0})
may be obtained easily by pushing one of the M5's to infinity.

\newpage

\fejezet{Interpretation}

The very fact that solutions exist with the boundary conditions (\ref{bc11}) immediately offers an interpretation
for our geometry. This is because the transverse coordinates $X_i$ close to one of the M5's
behave just as for a single M5 \cite{Basu:2004ed}. The M2's open up like a funnel into the M5, thus our geometry can be interpreted as
a finite length bi-funnel\footnote{Something that is not very useful in a household as opposed to a regular funnel, corresponding
to a stack of M2's ending on a single M5.}
opening up into M5-branes on both ends. On each end there is a self-dual string on the world volume of the M5
and hence our solution describes a kind of ``non-abelian self-dual string".
Again, this is very analogous to the D1-D3 brane system in type IIB string theory when
two D3-branes are placed at a finite distance from each other and D1's are suspended between them \cite{Diaconescu:1996rk}.
In this case the fact that two D3-branes are present as opposed to one results in a non-abelian
monopole as opposed to an abelian one.

\section{Conclusions}
\label{conc}

The solutions we have obtained for a stack of four membranes stretching between M5-branes may easily be embedded into
the $N\times N$ setting describing $N$ membranes. The ansatz $X_i(s) = ia_{ij}\,\gamma_j$ becomes $X_i(s)=ia_{ij}\,G_j$,
with $G_j$ being the $N\times N$ dimensional analogs of $\gamma_i$ with similar algebraic properties \cite{Basu:2004ed}.
In particular, the transverse coordinates $X_i$ to the self-dual string may be viewed to take values
in a fuzzy 3-sphere for which the matrices $G_i$ form a basis \cite{Balachandran:2005ew}.  
For general $N$, however, these are not the most general solutions, only a subset and it would be interesting to find a
complete description of the moduli space for any $N$.

Another interesting possibility would be studying periodically placed M5's and a periodic array of membranes stretching
between them. Such a configuration \dash involving an M5 with membranes ending on it from both sides \dash
would be suitable for dimensional reduction to string theory. The boundary conditions for the analogous
configuration of D3-branes with D1's ending on it from both sides are well-known in the context of
Nahm's transform for $SU(n)$ gauge theory.
However, we do not know what the boundary condition for the Basu-Harvey equation should be at the location of the M5
if membranes end on it from both sides. 

More generally, it would be most desirable to explore what aspects
of the Nahm transform and its properties are inherited by the generalized Nahm equation and what its implications are
for M-theory \dash if any.

Finally, we hope that our results might be used to gain some insight into the nature of the non-abelian tensor
theory that supposedly governs the dynamics of multiple M5-branes.

\section*{Appendix}

In this appendix we have collected the basic definitions and notions related to the quartic $f(x)=x^4+6px^2+4qx+r$,
see \cite{whittaker}. 

The roots $x_i$ of the quartic $f(x)$ can be expressed by square roots and roots of
the cubic $4u^3 - g_2 u - g_3$, where the elliptic invariants are
\bea
\label{g2g3}
g_2 = r + 3p^2\,,\qquad g_3 = pr - p^3-q^2\,.
\eea
Concretely, in terms of the roots $e_i$ of the cubic $4u^3-g_2 u - g_3 = 4(u-e_1)(u-e_2)(u-e_3)$ and $v_i = \sqrt{ - e_i - p }$ the roots of the quartic are
\bea
\label{quarticroots}
x_1 = \;\;v_1 + v_2 - v_3&\qquad& 
x_2 = \;\;v_1 - v_2 + v_3\nn\\
x_3 = -v_1 + v_2 + v_3&\qquad&
x_4 = -v_1 - v_2 - v_3\,.
\eea
Uniformization of the genus one curve $y^2=x^4+6px^2+4qx+r$ in terms of a parameter $s$ is
\bea
\label{uni}
x &=& x_1 + \frac{\frac{1}{4}f^\vesszo(x_1)}{\curly P(s)-\frac{1}{24}f^{\vesszo\vesszo}(x_1)}\\
y &=& \frac{-\frac{1}{4}f^\vesszo(x_1)\curly P^\vesszo(s)}{\left(\curly P(s)-\frac{1}{24}f^{\vesszo\vesszo}(x_1)\right)^2}\,,\nn
\eea
if $x_1$ is one of the roots, and where $\curly P(s) = \curly P(s\,;g_2,g_3)$ is the Weierstrass elliptic function
with invariants $g_2$ and $g_3$ given by (\ref{g2g3}). It satisfies the differential equation
\bea
\label{weierstrasseq}
\curly {P^\vesszo}^{\,2} = 4\,\curly P^3 - g_2 \curly P - g_3\,.
\eea
Thus uniformization solves our basic equation (\ref{df}), since $dx / ds = y$. We found it useful, however, to work with the Jacobi elliptic functions
defined by
\bea
\label{jacobidef}
z = \int_0^{\varphi(z|m)} \frac{dt}{\sqrt{1-m\sin^2t}}\,,\qquad \sn{z}{m} = \sin \varphi(z\,|\,m)\,,\qquad \cn{z}{m} = \cos\varphi(z\,|\,m)\,,
\eea
and $\dnn{z}{m}+m\,\snn{z}{m} = 1$. For $m=0$ we have $\sn{z}{0} = \sin(z)$, $\cn{z}{0}=\cos(z)$ and $\dn{z}{0} = 1$.
The following identities for the derivatives with respect to $z$ are useful for showing that (\ref{phi}-\ref{phisolutions})
is indeed a solution of (\ref{df}) and that there are the poles (\ref{pole}) at $s=\pm s_0$,
\bea
\label{id}
sn^\vesszo(z\,|\,m) &= &\cn{z}{m}\,\dn{z}{m}\nn\\
cn^\vesszo(z\,|\,m)& =& - \sn{z}{m}\,\dn{z}{m}\\
dn^\vesszo(z\,|\,m) &=& -m\,\sn{z}{m}\,\cn{z}{m}\,.\nn
\eea
Now the solution (\ref{phi}-\ref{phisolutions}) and the one above from uniformization
are seen to be equivalent due to the identity
\bea
\label{jw}
\curly P(z\,;g_2,g_3) = e_3 + \frac{e_1-e_3}{sn^2(z\sqrt{e_1-e_3}\,|\,\frac{e_2-e_3}{e_1-e_3})}\,.
\eea

We have also made use of the complete elliptic integral of the first kind, which may be 
defined by
\bea
\label{K}
K(m) = \int_0^{\pi/2} \frac{dt}{\sqrt{1-m \sin^2(t)}}\,.
\eea
The function $sn$ has the following periodicity property,
\bea
\label{jacobiprop}
sn(z+2qK(m)+2ipK(1-m)\,|\,m) = (-1)^q sn(z\,|\,m)\,,
\eea
for any integers $p$ and $p$. Thus we see that the requirement $s_0D<K(m)$ is needed for real $D$ and $m$
values for the function
$\phi(s)$, defined by (\ref{phi}), to have no pole on the interval $(-s_0,s_0)$ apart from the boundary.

\section*{Acknowledgements}

I am grateful to Jan de Boer, Jeff Harvey, Diederik Roest and Bella Schelpe for various enlightening comments,
remarks and discussion.

\end{document}